\documentstyle[12pt]{article}
\begin{document}
\begin{center}
{\bf{Chaplygin Gravitodynamics}}
\vskip 2cm
Rabin Banerjee\\
S.N.Bose National Centre for Basic Sciences,\\
JD Block, Sector III, Salt Lake City, Calcutta 700098, India
\vskip .5cm and \vskip .5cm

Subir Ghosh\\
Physics and Applied Mathematics Unit,\\
Indian Statistical Institute,\\
203 B. T. Road, Calcutta 700108, India.
\end{center}
\vskip 3cm
{\bf Abstract:}\\
We consider a new approach for gravity theory coupled to 
Chaplygin  matter in which the {\it{relativistic}} formulation of the latter
 is of crucial importance. We obtain a novel form of matter with
 dust like density $(\sim (volume)^{-1})$ and negative pressure. 
We explicitly show that our results are compatible with a relativistic generalization of the energy conservation principle, derived here.

\newpage
The equation of state,
\begin{equation}
p=-\frac{A}{\rho} \label{16}
\end{equation}
where $p$ and $\rho $ denote pressure and density respectively and $A$ being a parameter, was introduced way back by Chaplygin \cite{chap} as an effective model in computing the lifting force on a wing of an airplane. However, in modern cosmological context, the above model has turned out to be an interesting option in the 
Dark Energy models that initiate the acceleration of the
universe.  Being a perfect fluid system with negative pressure, this new role of Chaplygin gas  was anticipated in the work
of Jackiw et.al. \cite{jac} and others \cite{jac1,jac2,jac3} and was first explicitly introduced 
 in \cite{kam} as a model for Dark Energy.

However we observe that there exists 
a relativistic formulation of Chaplygin gas  that yields a generalized expression for pressure \cite{jac},
  \begin{equation}
p=-\frac{ma^2}{\sqrt{a^2+\rho ^2}}, \label{24a}
\end{equation}
 with $m$ and $a$ being constant parameters. In the non-relativistic limit, 
 $$p \approx
-\frac{ma^2}{\rho}+O(\frac{1}{\rho ^3})$$
 so that the relativistic equation of state (\ref{24a}) reduces to the conventional non-relativistic one (\ref{16}).

Our analysis using FRW gravity coupled with relativistic Chaplygin matter
shows that {\it{for all times}} the energy density will be of the dust-like form,
 \begin{equation}
\rho=\frac{b}{R^3}, \label{16b}
\end{equation}
where $b$ is an integration constant and $R$ is the scale factor. Although the form of density (\ref{16b}) is similar to a 
conventional dust it has a negative pressure, characteristic of the Chaplygin gas.  
This feature of the Chaplygin gas is shown to be compatible with a generalized energy conservation principle,
 pertaining to the relativistic Chaplygin dynamics. This is our main result.

The paper is divided into two major sections. In {\it{Section I}}, we present our analysis of (relativistic) Chaplygin cosmology, leading to (\ref{16b}). {\it{Section II}} discusses the points of distinction between our approach and the conventional one of \cite{kam,gen,gen1,gen2,gen3,gen4,gen5,gen6,gen7, astro,astro1,astro2,astro3,astro4}. Also we establish the compatibility between our result (\ref{16b}) with (\ref{24a}) by exploiting a relativistic generalization of energy conservation principle.
\vskip .5cm
{\it{Section I: Relativistic formulation of Chaplygin cosmology}}\\
\vskip .3cm
The starting  point is to consider the relativistic Chaplygin action in a flat spacetime, as  given by Jackiw et.al. \cite{jac},
\begin{equation}
{\cal A}_{Ch}=\int d^4x [-\dot \psi \rho -\sqrt{(\rho
^2+a^2)(m^2+ \partial^\mu \psi \partial_\mu \psi )}]. \label{1a}
\end{equation}
 We propose a natural generalization
 of the above action in FRW background by including the metric:
 \begin{equation}
{\cal A}_{Ch}=\int d^4x \sqrt{-g}[-\dot \psi \rho -\sqrt{(\rho
^2+a^2)(m^2+\vec \partial \psi ^2)}]. \label{1}
\end{equation}
We are using notations,
$$\partial \psi ^2\equiv g_{\mu\nu}\partial^\mu\psi\partial^\nu\psi~, ~
\vec\partial \psi ^2\equiv g_{ij}\partial^i\psi\partial^j\psi~,~
\dot \psi \equiv \frac{d\psi}{dt}.$$ In the flat background, it
has a  non-relativistic (large $m$)  limit for small $a^2$,
$$
{\cal L}_{Ch}=-\dot \psi \rho -m\rho
\sqrt{(1+\frac{a^2}{\rho^2})(1+\frac{\vec \partial
\psi^2}{m^2})}$$
\begin{equation}
\approx -(\rho\dot \psi_{NR} +\rho\frac{\vec \partial
\psi^{2}_{NR}}{2m}+\frac{ma^2}{2\rho}), \label{nrch}
\end{equation}
which is the standard Lagrangian of the Chaplygin fluid model with the
identification of the Chaplygin coupling $\lambda \equiv
\frac{ma^2}{2}$ and the mapping
\cite{jac} $\psi \equiv -mt +\psi_{NR}$. It is important to note that the matter (Chaplygin) action is in the Eulerian form in co-moving coordinates, that is compatible with the FRW equations.

The Chaplygin equations of motion from (\ref{1}) are,
\begin{equation}
\frac{d}{dt}(\sqrt{-g}\rho )+\partial^{j}[\frac{\sqrt{-g}\sqrt{\rho
^2+a^2}g_{ij}\partial^{i}\psi}{\sqrt{(m^2+\vec \partial \psi
^2)}}]=0~, \label{2}
\end{equation}

\begin{equation}
\dot \psi +\rho\sqrt{\frac{m^2+\vec \partial \psi ^2}{\rho
^2+a^2}}=0~. \label{3}
\end{equation}
One can eliminate $\rho$ by using (\ref{3}),
\begin{equation}
\rho =-\frac{a\dot \psi}{\sqrt{m^2+ \partial \psi ^2}}~. \label{4}
\end{equation}
This reduces the Chaplygin Lagrangian ${\cal L}_{Ch}$ (\ref{1}) to
the Born-Infeld Lagrangian in curved background when $\rho $ is
eliminated,
\begin{equation}
{\cal L}_{BI}=-a\sqrt{-g}\sqrt{m^2+ \partial \psi ^2}~. \label{5}
\end{equation}
The equation of motion for $\psi$ obtained from (\ref{5}) is,
\begin{equation}
\partial^{\mu}[\frac{\sqrt{g}g_{\mu\nu}\partial^{\nu}\psi}{\sqrt{m^2+ \partial \psi ^2}}]=0.
\label{6}
\end{equation}
This agrees with the  equation of motion for $\psi$ that one gets
from (\ref{2}) and (\ref{3}) after substituting $\rho$ from
(\ref{4}).

The utility of introducing the Born-Infeld form of the action is
the following. We wish to compute the Energy-Momentum Tensor (EMT)
$T_{\mu\nu}$ that is required in the Einstein equation. However, at
the same time we wish to include the $\rho$-field to make contact
with the Chaplygin cosmology. Now, the form of the action
containing $\rho$ in (\ref{1}) is not in a manifestly covariant form and so to get the
EMT in a covariant form from (\ref{1}) is awkward. On the other
hand, it is straightforward to get $T_{\mu\nu}$ from the
Born-Infeld model in (\ref{5}) but it does not contain $\rho$.
However, in $T_{\mu\nu}$ obtained from (\ref{5}), $\rho$ can be
reintroduced by exploiting (\ref{2})-(\ref{3}). We have already
checked the consistency of this procedure (see comment below
(\ref{6})). We will follow this route.

$T_{\mu\nu}$ obtained from (\ref{5}) is
\begin{equation}
T^{\mu\nu}=-a[\sqrt{m^2+ \partial \psi
^2}g^{\mu\nu}-\frac{\partial^{\mu}\psi\partial^{\nu}\psi}{\sqrt{m^2+
\partial \psi ^2}}]. \label{8}
\end{equation}
In particular we find,
\begin{equation}
T^{00}=-\frac{a}{\sqrt{m^2+ \partial \psi ^2}}[(m^2+ \partial \psi
^2)g^{00}-(\partial^{0}\psi )^2]
=\sqrt{(\rho ^2+a^2)(m^2+\vec \partial \psi ^2)}. \label{9}
\end{equation}
Note that this form of energy density agrees with the one that can
be read off  directly from the Chaplygin model (\ref{1}), the
latter being first order in time derivative. This again ensures
that the expression of the EMT from Born-Infeld and its subsequent
form involving $\rho$ is consistent. Obviously we are interested
in the  latter as this will involve the density $\rho$
directly.

The other components of the EMT are,
\begin{equation}
T^{i0}=T^{0i}=-\rho\partial^{i}\psi~, \label{10}
\end{equation}
\begin{equation}
T^{ij}=-a^2\sqrt{\frac{m^2+\vec \partial \psi ^2}{\rho
^2+a^2}}g^{ij}+\sqrt{\frac{\rho ^2+a^2}{m^2+\vec \partial \psi
^2}}\partial^{i}\psi\partial^{j}\psi~. \label{11}
\end{equation}

Notice that the $\rho ,\psi $-system, comprising of (\ref{1},\ref{2},\ref{3}) is not manifestly Lorentz or general coordinate invariant. Hence, as a necessary and sufficient condition for the Poincare invariance
of the theory as well as the consistent definition of the EMT (as
given in (\ref{9},\ref{10},\ref{11}) obtained from (\ref{8})), we have
explicitly checked the validity of the Schwinger conditions,
$$
\{T^{00}(\vec x),T^{00}(\vec y)\}=(T^{0i}(\vec x)+T^{0i}(\vec
y))\partial^{i}_{(x)}\delta (\vec x-\vec y),$$
\begin{equation}
\{T^{00}(\vec x),T^{0i}(\vec y)\}=(T^{00}(\vec x)+T^{00}(\vec
y))\partial^{i}_{(x)}\delta (\vec x-\vec y). \label{Sch}
\end{equation}
The above is computed by exploiting the (equal time) Poisson
Brackets,
\begin{equation}
\{\psi (\vec x),\rho (\vec y)\}= \delta (\vec x-\vec y)~,~~
\{\psi (\vec x),\psi (\vec y)\}=0~,~~\{\rho (\vec x),\rho (\vec
y)\}=0~, \label{PB}
\end{equation}
that can be read off from the symplectic structure of the
Lagrangian in (\ref{1}).

For completeness let us quickly derive the FRW equations from the
total action
\begin{equation}
{\cal{A}}={\cal{A}}_{Ch}+{\cal{A}}_{Gravity}. \label{action}
\end{equation}
 We consider a FRW
universe with the metric  $g_{\mu\nu}$ identified below,
\begin{equation}
-d\tau ^2=-dt^2+R^2(t)(\frac{dr^2}{1-kr^2}+r^2d\theta ^2+r^2sin
^2\theta d\phi ^2 ). \label{frw}
\end{equation}
In the Einstein equation
\begin{equation}
R_{\mu\nu}=-8\pi GS_{\mu\nu}
~,~S_{\mu\nu}=T_{\mu\nu}-\frac{1}{2}g_{\mu\nu}T^{\sigma}_{\sigma},
\label{7}
\end{equation}
the Einstein tensor $R_{\mu\nu}$ for FRW spacetime is,
\begin{equation}
R_{00}=\frac{3\ddot R}{R}~,~R_{ij}=-(R\ddot R+2\dot R^2
+2k)g_{ij}~,~R_{0i}=0. \label{12}
\end{equation}
In fact the elegance of our formalism will be manifest now since
{\it{everything}}, including $\rho =\rho (R)$ (given in (\ref{16b})) valid for Chaplygin
model, will be derived from the Einstein equation (\ref{7}) for
the present case. One should remember that in the analysis of
\cite{kam,gen,gen1,gen2,gen3,gen4,gen5,gen6,gen7} the analogus relation (\ref{16a}) was obtained separately, by incorporating the {\it{nonrelativistic}}
Chaplygin equation of state (\ref{16}) by hand.

We start by noting that in our case the $(0i)$ component of
Einstein equation is {\it not} vacuous. It yields
\begin{equation}
S_{0i}=-\rho\partial_{i}\psi =0. \label{20}
\end{equation}
Since $\rho \ne 0$, this in turn indicates $\partial_{i}\psi =0 $
ensuring homogeneity of $\psi $. This also makes $\rho$ uniform
since $\rho$ is related to $\psi$ by (\ref{4}). This restriction
considerably simplifies the rest of the source terms and using
(\ref{9},\ref{10},\ref{11}) we find,
\begin{equation}
S_{00}=\frac{m(\rho ^2-2a^2)}{2\sqrt{\rho
^2+a^2}}~,~S_{ij}=\frac{m(\rho ^2+2a^2)}{2\sqrt{\rho
^2+a^2}}g_{ij}~. \label{18}
\end{equation}
From the $(00)$ and $(ij)$ component of the Einstein equation,
after eliminating $\ddot R$, we find
\begin{equation}
\frac{\dot R^2}{R^2}=\frac{8\pi Gm}{3}\sqrt{\rho
^2+a^2}-\frac{k}{R^2}. \label{19}
\end{equation}
Notice that the matter contribution in (\ref{19}) is structurally
different from the corresponding equation in \cite{kam}.

The last task is to obtain $\rho$ as a function of $R$. For this
we put $\partial_{i}\psi =0 $ back in  the equation of motion
(\ref{2}), and get
\begin{equation}
 \frac{d}{dt}(\sqrt{-g}\rho )=0.
\label{20a}
\end{equation}
Let us try to obtain a solution of the differential equation
(\ref{20a}) directly. For FRW metric we compute
$\frac{d}{dt}(\sqrt{-g})=3\sqrt{-g}\frac{\dot R}{R}$ and
substitute this in (\ref{20a}) to find
\begin{equation}
\dot \rho =-3\rho \frac{\dot R}{R}, \label{b20}
\end{equation}
the solution of which is trivially obtained,
$$
\rho =\frac{b}{R^{3}},$$
with $b$ being an integration constant, as presented in (\ref{16b}). This is the central result of our paper.

Clearly, as stated before, the above form of density in (\ref{16b})   is completly different from the functional form of  $\rho (R)\sim \sqrt{A+BR^{-6}}$ mentioned in (\ref{16a}) that was  obtained in \cite{kam}.  Substituting
(\ref{16b}) in (\ref{19}) we obtain the evolution equation,
\begin{equation}
\frac{\dot R^2}{R^2}=\frac{8\pi
Gm}{3}\sqrt{a^2+\frac{b^{2}}{R^{6}}}-\frac{k}{R^2}. \label{21}
\end{equation}
Coincidentally this is the same equation that was obtained in
\cite{kam}.
\vskip .5cm
{\it{Section II: Discussions}}
\vskip .3cm

Let us now come to our relativistic scheme. As we have discussed in {\it{Section I}} in detail, in our formulation, the density function is derived naturally starting from the well defined  total relativistic action (\ref{action}). This function is induced solely from Chaplygin matter dynamics. It is derived from (\ref{2}) ( or (\ref{20a})) which are dynamical equations of motion of the Chaplygin matter. We emphasise that this feature of taking into account the Chaplygin matter dynamics is completely absent in the conventional analysis \cite{kam}.

Alternatively, one can also exploit the generic energy conservation principle,
\begin{equation}
d(energy~density\times volume)=-pressure\times d(volume)
\label{flaw}
\end{equation}
to derive the energy function, provided the proper relativistic generalizations of the terms appearing in (\ref{flaw}) are considered. In the present case the pressure of the relativistic Chaplygin gas is known to be (\ref{24a}). In fact, the general form of pressure is
obtained by identifying the Lagrangian with pressure in the
relativistic theory \cite{jac},
\begin{equation}
{p\equiv\cal{L}}_{BI}=-a\sqrt{m^2+\partial\psi
^2}=-a^2\sqrt{\frac{m^2+\vec\partial \psi ^2}{a^2+\rho
^2}}=-\frac{ma^2}{\sqrt{a^2+\rho ^2}}. \label{25}
\end{equation}
For simplicity we have dropped  $g$ from the Born-Infeld
Lagrangian (\ref{5}) and the last equality follows from the
uniformity of $\psi$ obtained in (\ref{20}). On the other hand, the relativistic energy density $T^{00}$ is derived from (\ref{9}):
\begin{equation}
T^{00} =m\sqrt{\rho ^2+a^2}.
\label{energy}
\end{equation}
Finally, a simple algebra will show that an identical form of density  (\ref{16b}), computed in {\it{Section I}} is recovered by using  the pressure function (\ref{24a}) or (\ref{25}) in  the relativistic energy conservation equation,  
\begin{equation}
d(T^{00}R^3)=-p~d(R^3),
\label{en}
\end{equation}
thereby proving our assertion.

We may mention that
in the conventional formulation of  Chaplygin cosmology \cite{kam}, the Universe evolves  smoothly from a dust
dominated to  de-Sitter model,  with the help of a
single fluid. This has been discussed in a  large number of theoretical \cite{gen,gen1,gen2,gen3,gen4,gen5,gen6,gen7} and
observational \cite{astro,astro1,astro2,astro3,astro4} studies.
The result follows from the particular form of energy density,
\begin{equation}
\rho=\sqrt{A+\frac{B}{R^6}}, \label{16a}
\end{equation}
with $A$ and $B$ being a constant parameter and an integration
constant respectively. The above relation is obtained by using the
equation of state for  Chaplygin gas \cite{kam},
in conjunction with energy conservation equation (\ref{flaw}). Subsequently, this form of density is substituted in the Friedman-Robertson-Walker (FRW) equation to generate the evolution equation for the scale factor.

  As we have shown this is not a unique way of interpreting Chaplygin cosmology since conclusions might differ if a
fully relativistic formulation is used. This indeed happens in the present case where, on the contrary, we find only a dust dominated Universe for all times.


\begin{thebibliography}{99}
\bibitem{chap}S.Chaplygin, Sci. Mem. Moscow Univ. Math. Phys. 21 (1904)1. 
\bibitem{jac} R. Jackiw, V.P. Nair, S.-Y. Pi, A.P. Polychronakos, J.Phys. A37 (2004) R327.
\bibitem{jac1}D. Bazeia, R. Jackiw, {\it{Nonlinear Realization of a Dynamical Poincare Symmetry by a Field-dependent Diffeomorphism}}, hep-th/9803165
\bibitem{jac2} D. Bazeia, Phys.Rev. D59 (1999) 085007.
\bibitem{jac3} M.Hassaine and P.Horvathy, Ann.Phys. (NY) 282 (2000)218.
\bibitem{kam}A.Kamenshchik, U.Moschella and V.Pasquier, Phys.Lett. B511 (2001)265.
\bibitem{gen}A.A.Sen, R.J.Scherrer, {\it{Generalizing the generalized Chaplygin gas}},
astro-ph/0507717.
\bibitem{gen1} M. K. Mak, T. Harko, Phys.Rev. D71 (2005)
104022.
\bibitem{gen2} V.Gorini, A.Kamenshchik, U.Moschella, V.Pasquier,
A.Starobinsky, {\it{Stability properties of some perfect fluid
cosmological models}}, astro-ph/0504576.
\bibitem{gen3} S. Capozziello, V.F.
Cardone, A. Troisi, Phys.Rev. D71 (2005) 043503.
\bibitem{gen4} L.P.Chimento,
R.Lazkoz, {\it{Duality extended Chaplygin cosmologies with a big
rip}}, astro-ph/0505254.
\bibitem{gen5} A.Frolov, L.Kofman and A.Starobinski,
Phys.Lett. B545 (2002)8.
\bibitem{gen6}  U.Debnath, A.Banerjee, S.Chakraborty,
Class.Quant.Grav. 21 (2004) 5609.
\bibitem{gen7} Varun Sahni, {\it{Dark Matter
and Dark Energy}}, astro-ph/0403324.
\bibitem{astro}O.Bertolami, P.T.Silva, {\it{Gamma-ray bursts as dark energy-matter
probes in the context of the generalized Chaplygin gas model}}, astro-ph/0507192.
\bibitem{astro1}
B. Mota, M. Makler, M.J. Reboucas, {\it{Detectability of Cosmic Topology in Generalized
Chaplygin Gas Models}}, astro-ph/0506499.
\bibitem{astro2}Dao-jun Liu, Xin-zhou Li,
Chin.Phys.Lett. 22 (2005) 1600.
\bibitem{astro3}  Zong-Hong Zhu
 Astron.Astrophys. 423 (2004) 421.
\bibitem{astro4} Yungui Gong,  JCAP 0503 (2005) 007.



\end{thebibliography}
\end{document}